\documentclass{PoS}
\usepackage{wrapfig}
\usepackage{cite}

\title{{\footnotesize DESY 13-020, DO-TH-13/04, SFB/CPP-13-011, LPN13-013}\\
ABM11 PDFs and the cross section benchmarks in NNLO}

\ShortTitle{ABM11 PDFs and benchmarking}

\author{\speaker{Sergey Alekhin}\\
        DESY, Platanenallee 6, D--15738 Zeuthen, Germany\\
Institute for High Energy Physics,142281 Protvino, Moscow region, Russia\\
        E-mail: \email{sergey.alekhin@desy.de}}

\author{Johannes Bl\"umlein\\
        DESY, Platanenallee 6, D--15738 Zeuthen, Germany\\
        E-mail: \email{johannes.bluemlein@desy.de}}

\author{Sven-Olaf Moch\\
II. Institut f\"ur Theoretische Physik, Universit\"at Hamburg 
    Luruper Chaussee 149, D-22761 Hamburg, Germany\\
        DESY, Platanenallee 6, D--15738 Zeuthen, Germany\\
        E-mail: \email{sven-olaf.moch@desy.de}}

\abstract{We report an updated version of the ABKM09 NNLO PDF fit, 
which includes the most recent HERA collider data on the inclusive cross 
sections and an improved treatment of the heavy-quark contribution to 
deep-inelastic scattering using advantages of the running-mass
definition for the heavy quarks. The ABM11 PDFs obtained from the updated 
fit are in a good agreement with the recent LHC data on the $W$- and 
$Z$-production within the experimental and PDF uncertainties. We also 
perform a determination of the strong coupling constant $\alpha_s$
in a variant of the ABM11 fit insensitive to the influence of the higher twist 
terms and find the value of $\alpha_s=0.1133(11)$ which is in good 
agreement with the nominal ABM11 one and our earlier determination.}

\FullConference{Loops and Legs in Quantum Field Theory - 11th DESY Workshop on Elementary Particle Physics,\\
		April 15-20, 2012\\
		Wernigerode, Germany}

\begin{document}

Modern phenomenology of high-energy hadronic interactions in 
great extent is driven by the precise calculations, which include 
higher order perturbative QCD corrections and require accurate knowledge 
of the non-perturbative input coming from the parton distribution 
functions (PDFs). 
In turn the latter are steadily improved due to 
the progress in the calculations of the perturbative corrections, 
which are commonly used to tune the PDFs, and  
refined measurements of those processes. The ABKM09 PDFs~\cite{Alekhin:2009ni}
were also recently updated to the ABM11 ones~\cite{Alekhin:2012ig}
taking advantage of more accurate 
deep-inelastic-scattering (DIS) data obtained by the HERA 
experiments~\cite{Aaron:2009aa,Collaboration:2010ry}
and improved theoretical treatment of the heavy-quark electro-production. 
This improvement is particularly important for the low-$x$ DIS data 
used in the analysis due to a significant 
\begin{wrapfigure}{r}{0.45\textwidth}
  \centering
  \includegraphics[width=0.4\textwidth]{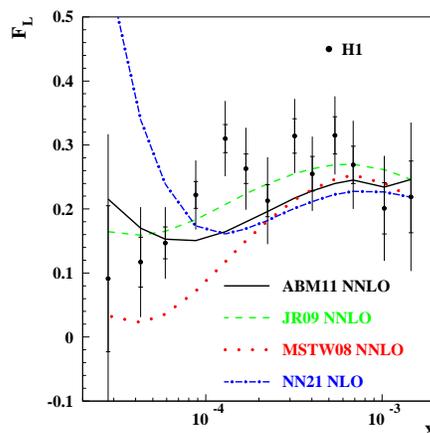}
  \caption{\small      The data on $F_L$ versus $x$ obtained by the H1
     collaboration~\cite{Collaboration:2010ry} 
     confronted with the 3-flavor scheme NNLO predictions based on the 
     different PDFs (solid line: this analysis, dashes: 
     JR09~\cite{JimenezDelgado:2008hf}, dots: MSTW~\cite{Martin:2009iq}). 
     The NLO predictions based on the 3-flavor NN21 
     PDFs~\cite{Ball:2011mu} are given for comparison (dashed dots). 
     The value of $Q^2$ for the data points and the curves in the plot 
     rises with $x$ in the range of $1.5 \div 45~{\rm GeV}^2$.}
  \label{fig:fl}
\end{wrapfigure}
contribution of the heavy quarks
in the inclusive cross sections. In the ABKM09 fit it is computed within 
the 3-flavor factorization scheme with the heavy quarks appearing in the 
final state. The ABM11 fit is also based on the 3-flavor scheme, however, 
the $\overline{\rm MS}$-mass definition is now used for the charm and 
bottom 
quarks, in contrast to the pole-mass definition employed in the ABKM09 fit. 
This approach provides improved convergence of the perturbative QCD 
calculations for the heavy-quark electro-production~\cite{Alekhin:2010sv}
which is crucial in view of the full NNLO corrections to 
this process are still missing. The quality of the ABM11 fit for the newly 
appended 
HERA data is quite good. The value of $\chi^2=537$ is obtained for the 
combined H1\&ZEUS sample~\cite{Aaron:2009aa}
including the neutral-current and the charged-current cross section data
with the total number of data points (NDP) equal to 486, while for the 
neutral-current data collected by the H1 experiment in the high-inelasticity 
run~\cite{Collaboration:2010ry} the value of $\chi^2/NDP=137/130$.
The latter data are particularly sensitive to the longitudinal structure 
function $F_L$ at small $x$, which is in turn sensitive to the small-$x$ 
gluon distribution. The 3-flavor scheme NNLO-predictions based on the 
ABM11, JR09, MSTW08, and NN21 PDFs are compared 
with the $F_L$ measurements of Ref.~\cite{Collaboration:2010ry} 
in Fig.~\ref{fig:fl}. While the ABM11 and JR09 analyses are based on the 
3-flavor treatment of the DIS, the MSTW08 and NN21 PDFs are obtained 
with different variants of the variable-flavor-number (VFN) scheme
with the charm quarks appearing in the initial state. Therefore in order 
to provide a consistent comparison the MSTW08 and NN21 predictions
are computed employing the complementary 3-flavor PDF sets provided by  
those groups. With such a unification the spread of 
the predictions in Fig.~\ref{fig:fl} displays the difference in
the corresponding PDFs, basically the small-$x$ gluon distributions. 
While the PDFs obtained within the 3-flavor scheme provide a better 
description of the data, the use of a modeling based on the VFN scheme 
apparently leads to departures at low values of $Q^2$.

In any case the small-$x$ gluons by 
different 
groups can evidently consolidate using these data. 
\begin{figure}[htb]
  \centering
  \includegraphics[width=0.45\textwidth]{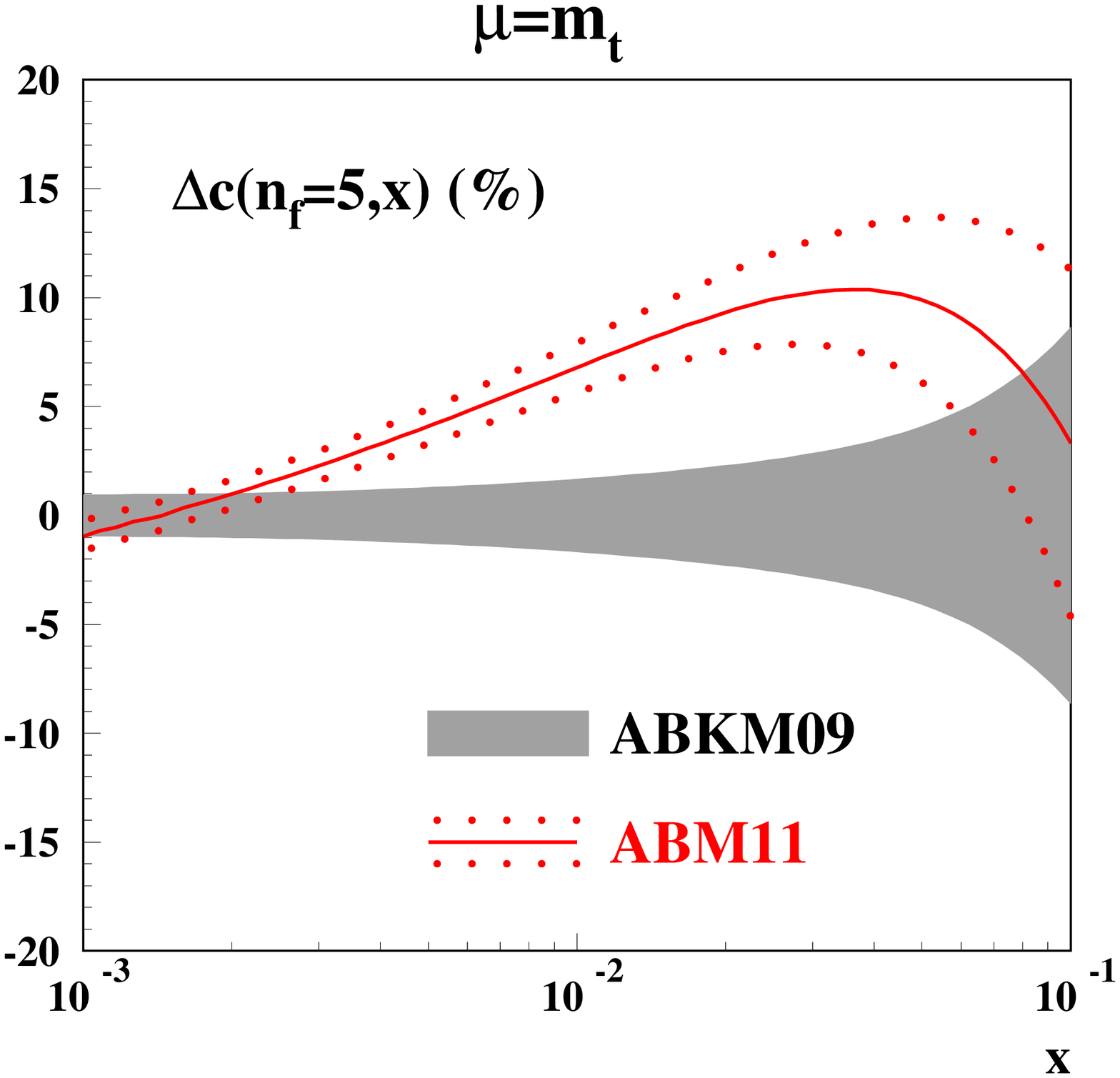}
  \includegraphics[width=0.45\textwidth]{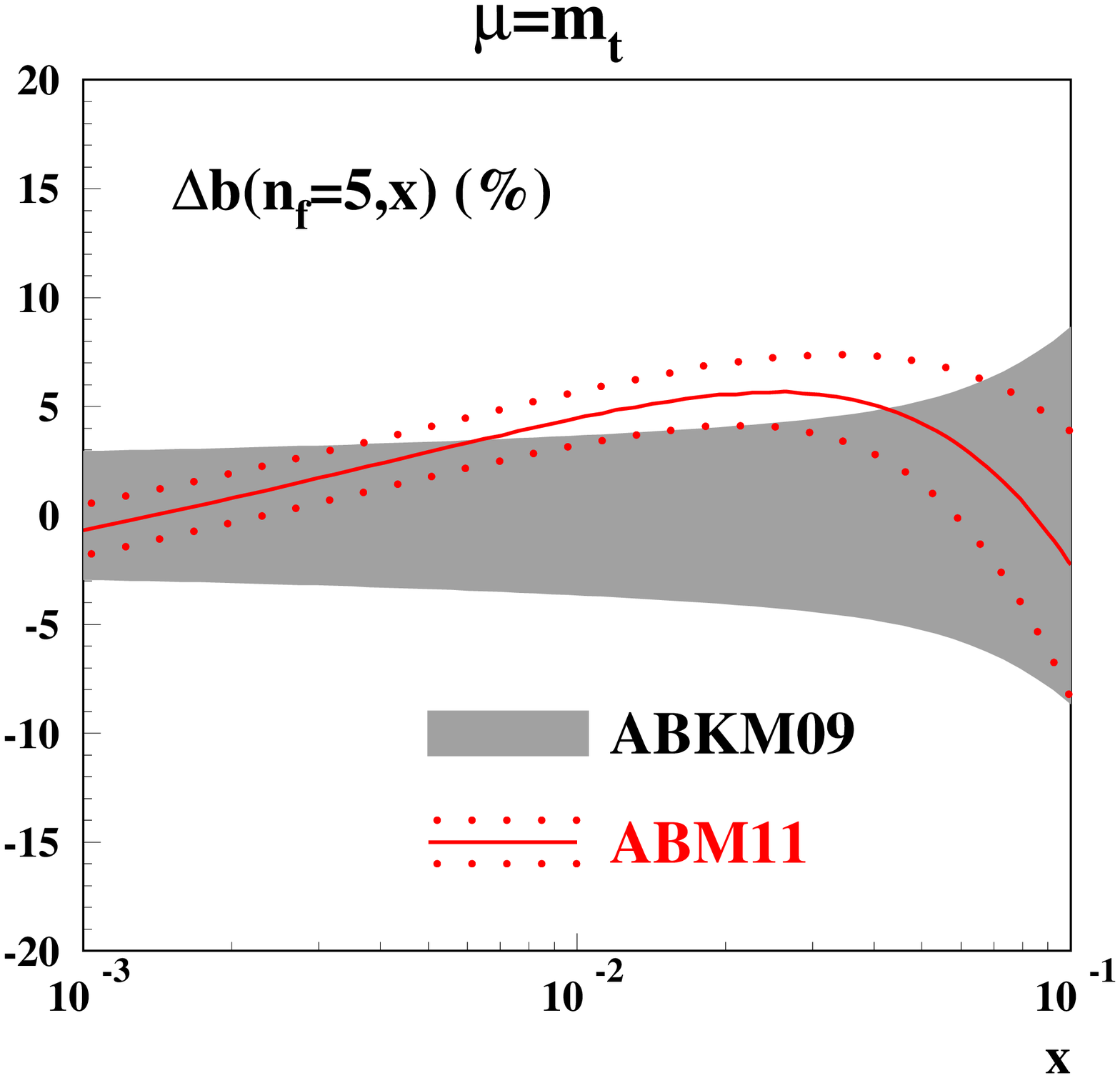}
  \caption{\small The charm- (left) and the bottom-quark (right) 
      PDFs obtained in the fit:
      The dotted (red) lines denote the $\pm 1 \sigma$ band of relative
      uncertainties (in percent) and the solid (red) line indicates the
      central prediction resulting from the fit with
      the running masses taken at the PDG values.
      For comparison the shaded (grey) area represents the 
      results of ABKM09~\cite{Alekhin:2009ni}.
}
  \label{fig:hq}
\end{figure}

The DIS HERA data are sensitive to the value of the charm-quark mass 
$m_c$. Its $\overline{\rm MS}$-value is obtained by $m_c(m_c) = 1.24\, \pm 
0.03 
(\rm{exp})\,^{+0.03}_{-0.02} (\rm{scale})\,^{+0.00}_{-0.07} (\rm{theory})$ GeV 
\cite{Alekhin:2012vu}
preferred by the inclusive HERA data of 
Refs.~\cite{Aaron:2009aa,Collaboration:2010ry} combined 
with the semi-inclusive data on open charm production~\cite{Abramowicz:1900rp}
is in good agreement with the world average~\cite{Beringer:1900zz}. 
This provides an additional argument in favor of the 3-flavor description 
of DIS. Furthermore, in the ABM11 analysis the accuracy of the small-$x$
4- and 5-flavor heavy-quark PDFs is improved 
as compared to the ABKM09 one due to the constraints on charm- and 
bottom-quark masses because the world average can be imposed, 
cf. Fig.~\ref{fig:hq}.
In contrast, the values of $m_c$ extracted from the HERA data 
within different 
variants of the VFN scheme demonstrate a wide spread
and, moreover, they are systematically lower than the ones obtained from the 
$e^+e^-$ data~\cite{Abramowicz:1900rp}.

\begin{figure}[h]
  \centering
  \includegraphics[width=\textwidth,height=9cm]{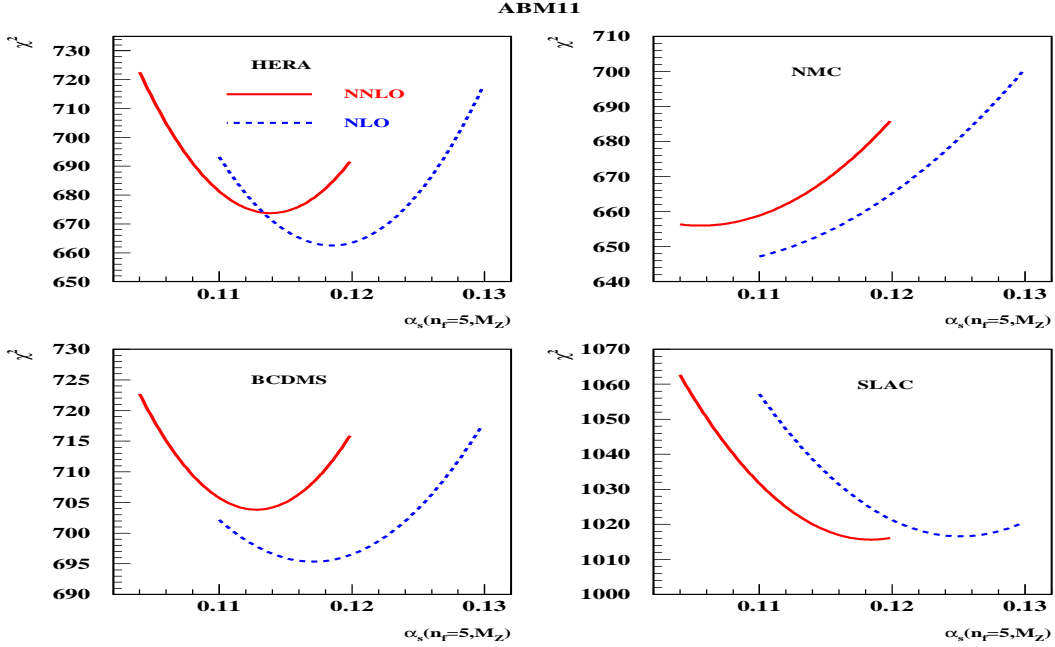}
  \caption{\small 
    The $\chi^2$-profile versus the value of $\alpha_s(M_Z)$, 
        for the separate data subsets,  
        all obtained in variants of the present analysis with the 
       value of $\alpha_s$ fixed 
    and all other parameters fitted
    (solid lines: NNLO fit, dashes: NLO fit).
}
  \label{fig:alp}
\end{figure}

The value of the strong coupling constant $\alpha_{\rm s}(M_{\rm Z})=0.1134(8)$ 
obtained in the NNLO ABM11 fit is in agreement with the earlier one 
of ABKM09. 
Meanwhile the NMC and SLAC DIS data sets prefer 
smaller and bigger value of $\alpha_{\rm s}$, 
respectively, cf. Fig.~\ref{fig:alp}.
Note that these two data sets are sensitive to the contribution of the 
higher twist terms~\cite{Alekhin:2012ig,Alekhin:2011ey}, while the HERA and 
BCDMS data do not suffer from this contribution due to larger typical 
values of 
the momentum transfer $Q^2$ involved. To avoid the influence of 
higher twist terms we perform a variant of the ABM11 fit with the SLAC 
and NMC data dropped and the higher twist terms set to zero. 
The value of $\alpha_{\rm s}(M_{\rm Z})=0.1133(11)$
obtained in this way is in a good agreement with the nominal value of 
ABM11
that demonstrates a consistent treatment of the higher twist terms in 
the present analysis. The NNLO ABM11 value of $\alpha_{\rm s}(M_{\rm Z})$ is 
essentially lower than the PDG world average~\cite{Beringer:1900zz}.  
On the other 
hand the ABM11 value of $\alpha_{\rm s}(M_{\rm Z})$ is in agreement 
with the recent NNLO determinations obtained from the analysis of thrust 
distributions measured in the $e^+e^-$ collisions with account of the power 
corrections~\cite{Abbate:2012jh,Gehrmann:2012sc} 
and also the lattice calculation of the QCD static 
energy~\cite{Bazavov:2012ka}.

\begin{figure}[t]
  \centering
  \includegraphics[width=0.9\textwidth,height=7cm]{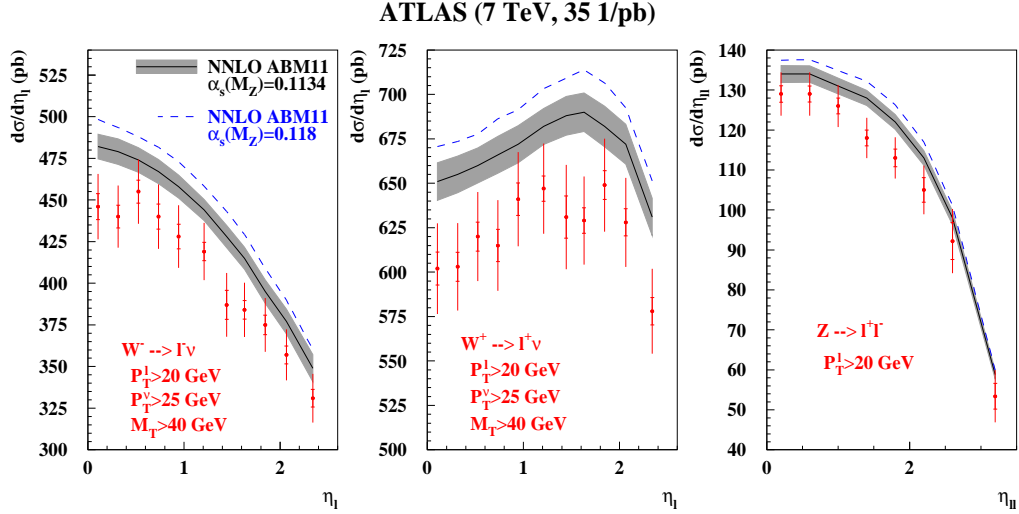}
\caption{\small The ATLAS data of Ref.~\cite{Aad:2011dm} on the 
rapidity distribution of the charged leptons 
$l^-$ (left panel), $l^+$ (central panel), and
$Z$-boson (right panel) in comparison with 
the NNLO predictions computed with the ABM11 PDFs taking
into account the uncertainties due to PDFs (shaded area). The dashed curves 
display the ABM11 predictions obtained with the value of 
$\alpha_s(M_Z)=0.118$.
}
\label{fig:atlas}
\end{figure}

Benchmarks for the rates of key-processes at Tevatron and the LHC were given in
\cite{Alekhin:2009ni,Alekhin:2012ig,Alekhin:2010dd}.
For the first LHC data on the Drell-Yan differential cross 
sections~\cite{Aad:2011dm,Chatrchyan:2012xt,Aaij:2012vn,lhcbe}
we observe good agreement with the predictions based on the ABM11 PDFs.
These data were
obtained by detecting the leptonic decays of $W$- and $Z$-bosons
with the charged lepton pseudo-rapidities 
in the range of $-2.5\div 2.5$, $-2.4\div 2.4$, and $2\div 4.5$
for the ATLAS, CMS, and LHCb experiments, respectively. The data
on the $Z$-boson pseudo-rapidities and on the charged lepton 
pseudo-rapidities for the case of $W$-boson production are confronted 
to the NNLO predictions obtained with ABM11 PDFs 
in Figs.~\ref{fig:atlas}--%
\ref{fig:lhcbe}.
The central values of the predictions are obtained using a fully 
exclusive code DYNNLO~1.3~\cite{Catani:2009sm}. However, the uncertainties 
due to PDFs are computed with another fully exclusive code 
FEWZ~3.1~\cite{Li:2012wn}, which allows to estimate the variation 
of the cross section with the PDFs without accumulating the 
Monte-Carlo integration errors in the cross section differences. 
The typical accuracy of the DYNNLO results displayed in 
Figs.~\ref{fig:atlas}--%
\ref{fig:lhcbe} is better 
than 1\% and it does not exceed the data errors.

Agreement between the data and the predictions is quantified by 
the following $\chi^2$ functional 
\begin{equation}
\chi^2=\sum_{i,j} (y_i-f_i^{(0)})[C^{-1}]_{ij}(y_j-f_j^{(0)}),
\label{eq:chi}
\end{equation}
with the covariance matrix 
\begin{equation}
C_{ij}=C_{ij}^{exp} + \sum_{k=1}^{28}(f^{(k)}_i-f^{(0)}_i)(f^{(k)}_j-f^{(0)}_j). 
\label{eq:covm}
\end{equation}
The first term in Eq.~(\ref{eq:covm}) corresponds to the
covariance matrix describing the errors in the 
experimental data, while the second term appears due to 
the PDF uncertainties. The latter contribute to the covariance matrix 
by virtue of the differences between 
predictions obtained for the central ABM11 PDF set, $f_j^{(0)}$, and 
the ones for 
each of the 28 ABM11 PDF sets representing the PDF uncertainties, $f_j^{(k)}$. 
The experimental covariance matrix for the ATLAS data of 
Ref.~\cite{Aad:2011dm} is computed by
\begin{equation}
C_{ij}^{exp}=\delta_{ij}\sigma_i^2 + f^{(0)}_i f^{(0)}_j \sum_{k=1}^{31}s_i^ks_j^k,
\label{eq:covexp}
\end{equation}
where $\sigma_i$ are the statistical errors in the data combined in 
quadrature with the uncorrelated errors, $s_i^k$ are the correlated 
systematic uncertainties representing 31 independent sources including 
the normalization, and $\delta_{ij}$ is the Kronecker symbol. In view of the small
background for the $W$- and $Z$-production signal all systematic errors
are considered as multiplicative. Therefore, 
they are weighted with the theoretical predictions $f^{(0)}_i$.
The experimental covariance matrices 
for the CMS and LHCb data of Refs.~\cite{Chatrchyan:2012xt,Aaij:2012vn,lhcbe} 
are available in publications and they are directly employed in 
Eq.~(\ref{eq:covm}) after re-weighting by the theoretical predictions
similarly to Eq.~(\ref{eq:covexp}) with the normalization uncertainty 
taken into account in the same way as for the ATLAS data. 
\begin{figure}[h]
  \centering
  \includegraphics[width=0.9\textwidth,height=10cm]{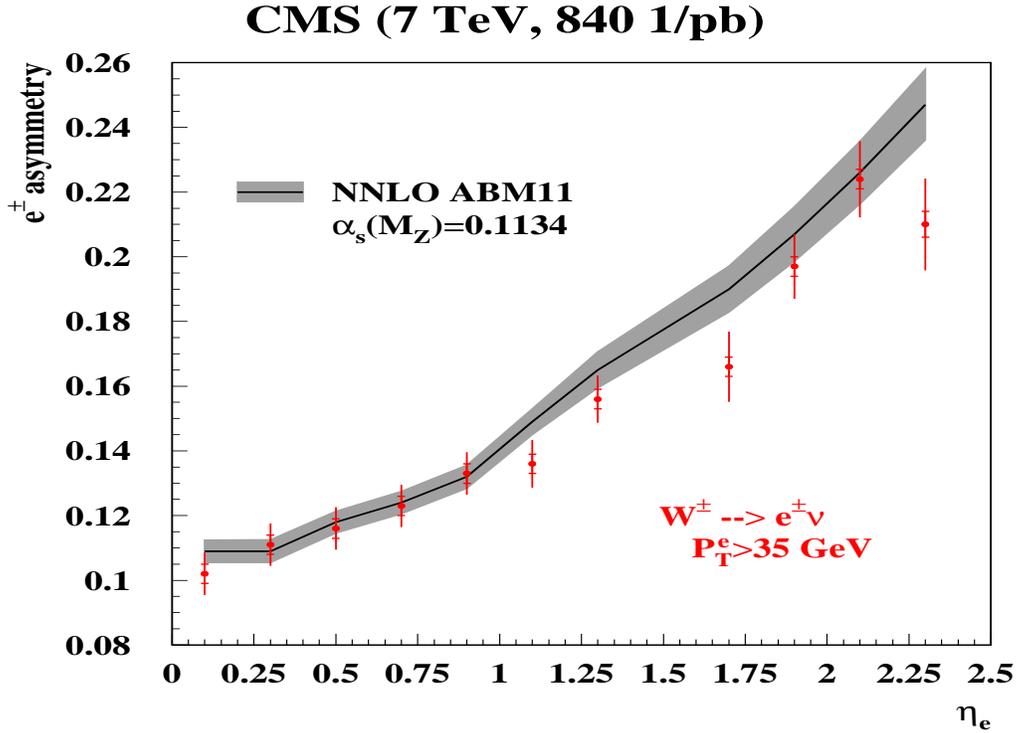}
  \caption{\small The same as Fig.~\protect\ref{fig:atlas} for the 
CMS data of Ref.~\cite{Chatrchyan:2012xt} on the electron asymmetry 
rapidity distribution. The curve corresponding to the ABM11 predictions 
obtained with the value of $\alpha_s(M_Z)=0.118$ coincides with the nominal one.
}
\label{fig:cms}
\end{figure}

The values of $\chi^2$ for different LHC data samples are summarized in 
Table~1. Due to an inconsistency in the covariance matrix for the LHCb 
$Z$-boson~data \cite{lhcbcov}
quoted in Ref.~\cite{Aaij:2012vn} the $Z$-boson data are
not employed in the calculation of $\chi^2$ for this LHCb sample.
In general the values of $\chi^2/NDP$ in 
Table~1 are comparable with 1 within possible 
statistical fluctuations. 
The ATLAS data~\cite{Aad:2011dm}
systematically undershoot the predictions, although the
statistical discrepancy of the disagreement is not significant in view 
of still relatively large experimental uncertainties.
Moreover, it is worth noting that the recent high-luminosity
LHCb data~\cite{lhcbe} on the $Z$-boson production 
are in much better agreement with the ABM11 predictions. In particular 
this is also observed for the rapidity range overlapping with the 
corresponding ATLAS data that
may indicate a potential inconsistency between the 
ATLAS and LHCb normalization. The ABM11 predictions
demonstrate a certain tension with 
the earlier LHCb data~\cite{Aaij:2012vn} on the $W^+$-boson production 
at moderate rapidity, however, the overall agreement is satisfactory. 
If, nonetheless, the discrepancies with the 
ATLAS and LHCb data being observed are
confirmed with better statistical significance, the ABM11 quark distributions 
at $x\sim 0.01$ would have to be lowered in order to accommodate 
those data in a fit. Meanwhile, the charged-lepton asymmetry which is 
less 
sensitive to the experimental uncertainties is in a good agreement 
with the ABM11 predictions both for the ATLAS and CMS data. 

\begin{figure}[htb]
  \centering
  \includegraphics[width=0.9\textwidth,height=7cm]{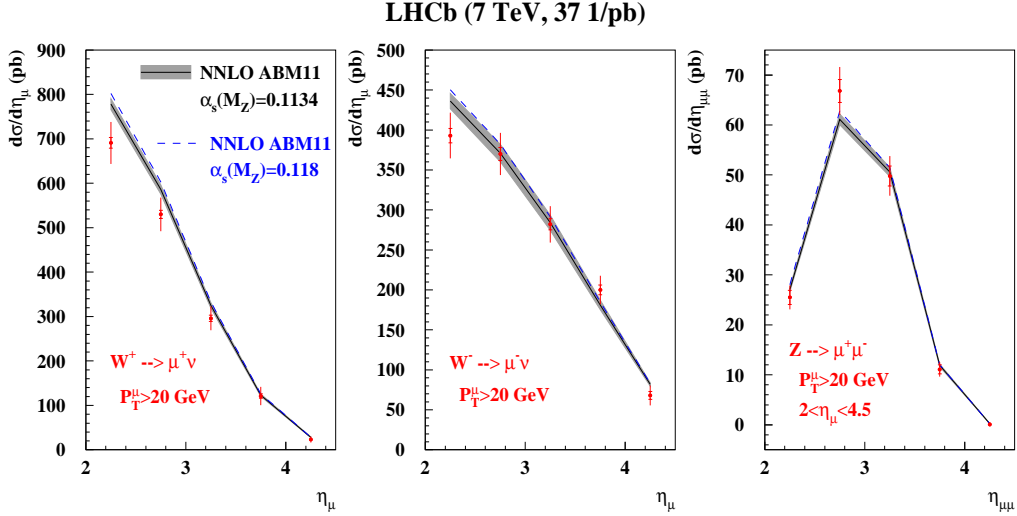}
  \caption{\small The same as Fig.~\protect\ref{fig:atlas} for the 
LHCb data of Ref.~\cite{Aaij:2012vn} on the 
rapidity distribution of $\mu^+$ (left panel), $\mu^-$ (central panel), and
$Z$-boson (right panel).}
\label{fig:lhcb}
\end{figure}

\begin{figure}[htb]
  \centering
  \includegraphics[width=0.9\textwidth,height=10cm]{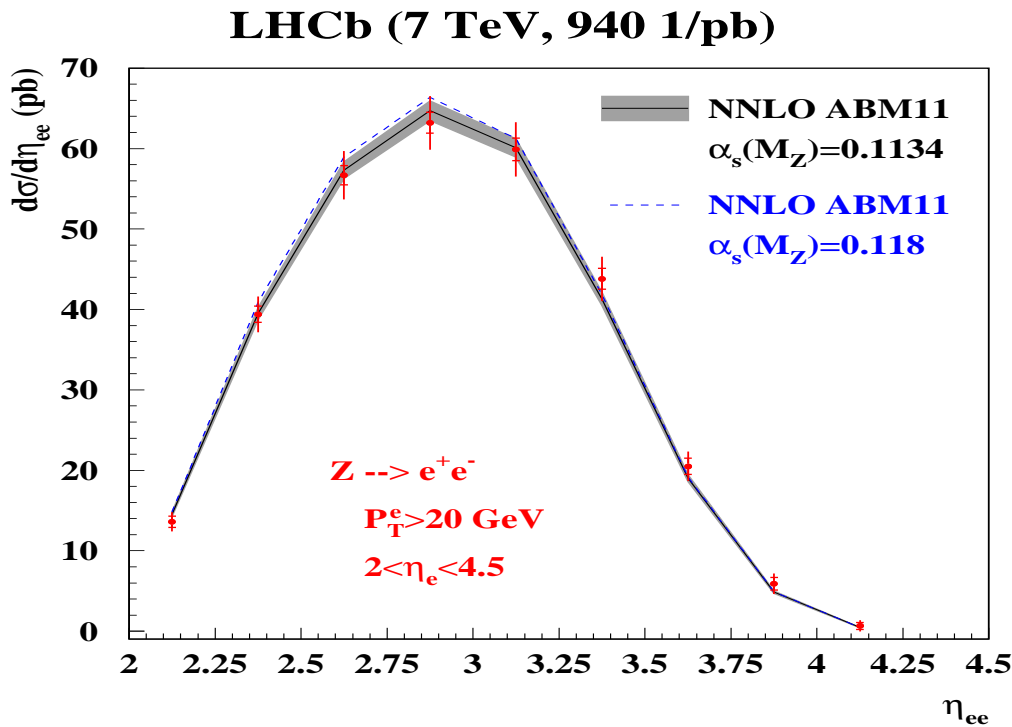}
  \caption{\small The same as Fig.~\protect\ref{fig:atlas} for the 
LHCb data of Ref.~\cite{lhcbe} on the 
$Z$-boson rapidity distribution.}
\label{fig:lhcbe}
\end{figure}

\begin{table}[th!]
\renewcommand{\arraystretch}{1.3}
\begin{center}                   
{\small                          
\begin{tabular}{|c|c|c|c|c|}   
\hline                           
{Experiment}                      
&ATLAS~\cite{Aad:2011dm}                         
&{CMS~\cite{Chatrchyan:2012xt}}  
&{LHCb~\cite{Aaij:2012vn}}
&{LHCb~\cite{lhcbe}}                            
\\                                                        
\hline
{Final states}                                                   
& $W^+\rightarrow l^+\nu$
& $W^+\rightarrow e^+\nu$
&$W^+\rightarrow \mu^+\nu$
& $Z\rightarrow e^+e^-$                                                        
\\                                                        
& $W^-\rightarrow l^-\nu$
&$W^-\rightarrow e^-\nu$
&$W^-\rightarrow \mu^-\nu$
&                                                         
\\                                                        
& $Z\rightarrow l^+l^-$
&                                                         
&                                                         
&                                                         
\\                                                        
\hline                                                    
{Luminosity (1/pb)}                      
&35                         
&840  
&{37}
&{940}                            
\\                                                        
\hline                                                    
$NDP$
&30                      
&11  
&10
&9                            
\\                                                        
\hline
 $\chi^2$
 &$35.7(7.7)$
 &$10.6(4.7)$
 &13.1(4.5)
 &11.3(4.2)
\\
\hline                                          
\end{tabular}
}
\caption{\small The value of $\chi^2$ obtained for different samples of 
the Drell-Yan LHC data with the NNLO ABM11 PDFs.
The figures in parenthesis give one standard deviation of 
$\chi^2$ equal to $\sqrt{2NDP}$.}
\end{center}
\label{tab:chi2}
\end{table}

The LHC Drell-Yan data were confronted to the predictions 
based on the 
different PDFs including ABM11 ones in Ref.~\cite{Ball:2012wy} earlier. 
In contrast 
with our analysis the NNLO predictions of Ref.~\cite{Ball:2012wy} are 
obtained using the NLO MCFM code~\cite{mcfm} supplemented by the 
NNLO K-factors which were computed separately with the DYNNLO code. 
Furthermore, the benchmarking of Ref.~\cite{Ball:2012wy} is performed using 
the PDFs corresponding to the value of $\alpha_s(M_Z)=0.118$. This 
value is disfavored by the ABM11 fit. Nonetheless, in order to allow a 
consistent comparison with the benchmarking of Ref.~\cite{Ball:2012wy},
we also provide the ABM11 prediction corresponding to the value of
$\alpha_s(M_Z)=0.118$. The predictions obtained in 
this way are compared with the data and the nominal ABM11 predictions in 
Figs.~\ref{fig:atlas},\ref{fig:cms},\ref{fig:lhcb},\ref{fig:lhcbe}. 
The charged-lepton asymmetry does not depend on the $\alpha_s$ value, 
while
the shift in the $W$ and $Z$ absolute cross-sections obtained with
this variation of $\alpha_s$ is of the order of the
uncertainties due to PDFs. 
{The value of $\chi^2=14.3$ obtained for the LHCb 
data~\cite{Aaij:2012vn} with the enhanced value of $\alpha_s$ 
is within the statistically allowed range.
Meanwhile the value of $\chi^2$ obtained in this case 
for the ATLAS data~\cite{Aad:2011dm} 
is 41.5, i.e. slightly above the $1\sigma$ statistical margin.
This is obviously related to the offset of the nominal ABM11 
predictions w.r.t. the ATLAS data. Therefore the final conclusion 
about the significance of this excess can be drawn only after 
the normalization discrepancy for ATLAS and LHCb is clarified.}
Anyway, both values are significantly smaller than those
quoted for the ABM11 benchmarking in Ref.~\cite{Ball:2012wy}.
{In particular this happens due to the PDF uncertainties were not 
taken into account in the statistical analysis of Ref.~\cite{Ball:2012wy}.
The central values for the ABM11 predictions obtained in that benchmarking 
also do not agree with ours in places, cf. 
Fig.~12 in Ref.~\cite{Ball:2012wy} and Fig.~\ref{fig:atlas} in the present 
paper~\footnote{To facilitate further clarification of the discrepancy 
the tabulated NNLO ABM11 predictions for the available ATLAS, 
CMS, and LHCb Drell-Yan data are attached 
to the arXiv version of these Proceedings.}.}
The recent LHCb data on the $Z\rightarrow e^+e^-$ production were not 
included into the benchmarking of Ref.~\cite{Ball:2012wy}, 
however, they are compared 
with the NNLO predictions based on the CT10~\cite{Nadolsky:2012ia},
MSTW09~\cite{Martin:2009iq}, and 
NN21~\cite{Ball:2011uy} PDFs in Fig.~4 of Ref.~\cite{lhcbe}. These PDFs 
provide consistent results. Nonetheless all three  
undershoot the LHCb $Z$-boson data at large rapidity. 
{To quantify this disagreement we check $\chi^2$ 
for these data taking NNLO MSTW08 predictions with account of the 
PDF uncertainties 
and find the value of 26.2, more than twice bigger than ours, cf. Table~1.}

In summary, we present the updated version of the NNLO PDF fit based 
on the combination of the DIS and Drell-Yan data. The update includes the 
most recent HERA collider data on the inclusive neutral-current and 
charged-current cross sections. From the theory side we improved the 
treatment of 
the heavy-quark contribution to DIS using advantages of the running-mass
definition. The ABM11 PDFs obtained from the updated fit 
are in a good agreement with the recent LHC 
data on the $W$- and $Z$-production within the experimental and PDF
uncertainties. Our results do not confirm the benchmarking of the 
ABM11 PDFs published elsewhere~\cite{Ball:2012wy}.
We also perform a determination of the $\alpha_s$ value in a
variant of the ABM11 fit insensitive to the higher twist terms
and find a value, which is in very good agreement with the nominal one
and our earlier determination of $\alpha_s$.

{\bf Acknowledgments.} We acknowledge fruitful discussions concerning 
the Drell-Yan LHC data with Uta Klein and Katharina M\"{u}ller. We are 
also grateful to Juan Rojo for clarification of some details of 
Ref.~\cite{Ball:2012wy}.

\end{document}